\documentclass[sigconf]{acmart}

\usepackage{booktabs}
\usepackage{balance}
\usepackage{url}
\usepackage{color}
\usepackage{multirow}
\usepackage{float}
\usepackage{mathtools}

\usepackage{amsthm}

\usepackage[ruled,vlined,linesnumbered]{algorithm2e}

\newtheorem{defn}{Definition}

\begin{document}

\settopmatter{printacmref=false} 

\copyrightyear{2018} 
\acmYear{2018}  
\setcopyright{acmcopyright}
\acmConference[SIGIR '18]{The 41st International ACM SIGIR Conference on Research
and Development in Information Retrieval}{July      8--12, 2018}{Ann Arbor, MI, USA}
\acmBooktitle{SIGIR '18: The 41st International ACM SIGIR Conference on Research and
Development in Information Retrieval, July      8--12, 2018, Ann Arbor, MI, USA}
\acmPrice{15.00}
\acmDOI{10.1145/3209978.3210063}
\acmISBN{978-1-4503-5657-2/18/07}


\title[Equity of Attention: Amortizing Individual Fairness in Rankings]{Equity of Attention: \\Amortizing Individual Fairness in Rankings}

\newcommand{\norm}[1]{\lVert#1\rVert}

\author{Asia J. Biega}
\affiliation{
 \institution{Max Planck Institute for Informatics}
 \city{Saarland Informatics Campus}
}
\email{jbiega@mpi-inf.mpg.de}

\author{Krishna P. Gummadi}
\affiliation{
 \institution{MPI-SWS}
 \city{Saarland Informatics Campus}
}
\email{gummadi@mpi-sws.org}

\author{Gerhard Weikum}
\affiliation{
 \institution{Max Planck Institute for Informatics}
 \city{Saarland Informatics Campus}
}
\email{weikum@mpi-inf.mpg.de}

\newcommand{\squishlist}{
   \begin{list}{$\bullet$}
    { \setlength{\itemsep}{0pt}      \setlength{\parsep}{3pt}
      \setlength{\topsep}{3pt}       \setlength{\partopsep}{0pt}
      \setlength{\leftmargin}{1.5em} \setlength{\labelwidth}{1em}
      \setlength{\labelsep}{0.5em} } }
\newcommand{\squishlisttwo}{
   \begin{list}{$\bullet$}c
    { \setlength{\itemsep}{0pt}    \setlength{\parsep}{0pt}
      \setlength{\topsep}{0pt}     \setlength{\partopsep}{0pt}
      \setlength{\leftmargin}{0.9em} \setlength{\labelwidth}{0.5em}
      \setlength{\labelsep}{0.5em} } }

\newcommand{\squishend}{
    \end{list} 
}

\begin{abstract}
Rankings of people and items are at the heart of selection-making, match-making, and recommender systems, ranging from employment sites to sharing economy platforms. As ranking positions influence the amount of attention the ranked subjects receive, biases in rankings can lead to unfair distribution of opportunities and resources such as jobs or income.

This paper proposes new measures and mechanisms to quantify and mitigate unfairness from a bias inherent to all rankings, namely, the {\it position bias} which leads to disproportionately less attention being paid to low-ranked subjects. Our approach differs from recent fair ranking approaches in two important ways. First, existing works measure unfairness at the level of subject {\it groups} while our measures capture unfairness at the level of {\it individual} subjects, and as such subsume group unfairness. Second, as no single ranking can achieve individual attention fairness, we propose a novel mechanism that achieves {\it amortized fairness}, where attention accumulated across a series of rankings is proportional to accumulated relevance.

We formulate the challenge of achieving amortized individual fairness subject to constraints on ranking quality as an online optimization problem and show that it can be solved as an integer linear program. Our experimental evaluation reveals that unfair attention distribution in rankings can be substantial, and demonstrates that our method can improve individual fairness while retaining high ranking quality.

\end{abstract}


\maketitle

\begin{section}{Introduction}

\noindent{\textbf{Motivation and Problem. }}
Rankings of subjects like people, hotels, or songs are at the
heart of selection, matchmaking and recommender systems. 
Such systems are in use on a variety of platforms that affect different aspects of life -- 
from entertainment and dating all the way to employment and income.
Notable examples of platforms with a tangible impact on people's livelihood include two-sided sharing economy websites, 
such as Airbnb or Uber, or human-resource matchmaking platforms, such as LinkedIn or TaskRabbit.
The ongoing migration to online markets and the growing dependence 
of many users on these platforms in securing an income have spurred 
investigations into the issues of bias, discrimination and fairness in the platforms' mechanisms \cite{calo2017taking,levy2017designing}.

One aspect in particular has evaded scrutiny thus far -- 
to be successful on these platforms, ranked subjects need to gain the {\em attention} of searchers.
Since exposure on the platform is a prerequisite for attention, subjects
have a strong desire to be highly ranked.
However, when inspecting ranked results, searchers are susceptible to {\em position bias},
which makes them pay most of their attention to the top-ranked subjects.
As a result, lower-ranked subjects often receive disproportionately less attention 
than they deserve according to the ranking relevance.
Position bias has been studied in information retrieval in scenarios where subjects are documents 
such as web pages (e.g., \cite{craswell2008experimental,chuklin2015click}).
It has been shown that top-ranked documents receive most clicks
often irrespective of their actual relevance
\cite{JoachimsRadlinski2007}.

Systemic correction for the bias becomes important when ranking positions potentially translate to financial gains or losses.
This is the case when ranking people on platforms like LinkedIn or Uber, products on platforms like Amazon,
or creative works on platforms like Spotify.
For example, cumulating the exposure on a subset of drivers in ride-hailing platforms might lead to economic starvation of others, 
while low-ranked artists on music platforms might not get their deserved chance of earning royalties.

Observing that attention is influenced by a human perception bias, 
while relevance is not, uncovers a fundamental problem: there necessarily exists a discrepancy between the 
attention that subjects receive at their respective ranks and
their relevance in a given search task.
For example, attention could decrease geometrically, whereas
relevance scores may decrease linearly as the rank decreases.
If a ranking is displayed unchanged to many searchers over time,
the lower-ranked subjects might be systematically 
and repeatedly disadvantaged in terms of the attention they receive.

\noindent{\bf Problem Statement.}
A vast body of ranking models literature has focused on aligning system relevance scores 
with the true relevance of ranked subjects, and in this paper we assume the two are proportional. 
What we focus on instead is the relation between relevance and attention.
Since relevance can be thought of as a proxy for
worthiness in the context of a given search task, the attention a subject receives from searchers 
should ideally be proportional to her relevance.
In economics and psychology, a similar idea of proportionality exists under the name of equity \cite{walster1973new}
and is employed as a fairness principle in the context of distributive justice \cite{greenberg1987taxonomy}.
Thus, in this paper, we make a translational normative claim and argue for {\em equity of attention} in rankings.

Operationally, the problem we address in this paper is to devise measures
and mechanism which ensure that, for all subjects in the system, the received attention
approximately equals the deserved attention, while preserving ranking quality.
For a single ranking this goal is infeasible, 
since attention is influenced by the position bias, while relevance is not.
Therefore, our approach looks at a series of rankings and aims at
measures of {\em amortized fairness}.

\noindent{\textbf{State of the Art and Limitations. }}
Fairness has become a major
concern for decision-making systems based on
machine learning methods (see, e.g., \cite{FATconference,DBLP:journals/ker/RomeiR14}).
Various notions of {\em group fairness} have been investigated 
\cite{DBLP:conf/pkdd/KamishimaAAS12,DBLP:conf/kdd/PedreschiRT08,Feldman:KDD2015,DBLP:conf/nips/HardtPNS16, DBLP:conf/www/ZafarVGG17},
with the goal of making sure 
that protected attributes such as gender or race do not influence algorithmic decisions.
Fair classifiers are then trained to maximize
accuracy subject to group fairness constraints.
These approaches, however, do not distinguish between different subjects from within a group.
The notion of {\em individual fairness} \cite{dwork2012fairness,DBLP:conf/icml/ZemelWSPD13,Kearns:ICML2017}
aims at treating each individual fairly by requiring that subjects
who are similar to each other receive similar decision outcomes.
For instance, the concept of {\it meritocratic fairness} 
requires that less qualified candidates are almost never
preferred over more qualified ones when selecting 
candidates from a set of diverse populations. Relevance-based
rankings, where more relevant subjects are ranked higher than less
relevant ones, also satisfy meritocratic fairness.
A stronger fairness concept, however, is needed 
for rankings to be a means of distributive justice.

Prior work on {\em fair rankings} is scarce and includes approaches 
that perturb results to guarantee various types of
group fairness. This goal is achieved by techniques similar to those for
ranking result diversification  \cite{celis2017ranking, yang2016measuring, zehlike2017fair},
or by granting equal ranking exposure to groups \cite{SinghJoachims2017}.
Individual fairness is inherently beyond the scope of group-based perturbation.

\noindent{\textbf{Approach and Contribution. }}
Our approach in this paper differs from the prior work in two
major ways. First, the measures introduced here 
capture fairness at the level of {\it individual}
subjects, and subsume group fairness as a special case. 
Second, as no single 
ranking can guarantee fair attention to every subject,
we devise a novel mechanism that ensures {\it amortized fairness},
where attention is fairly distributed across a series of rankings.

For an intuitive example,
consider a ranking where all the relevance scores are almost the same.
Such tiny differences in relevance will push subjects apart in the display of the results,
leading to a considerable difference in the attention received from searchers.
To compensate for the position bias, we can reorder the subjects in consecutive rankings
so that everyone who is highly relevant is displayed at the top every now and then.

Our goal is not just to balance attention, but to keep it proportional to relevance for all subjects
while preserving ranking quality.
To this end, we permute subjects in each ranking so as to improve fairness subject to constraints on quality loss.
We cast this approach to an online optimization problem, formalizing
it as an integer linear program (ILP). We moreover devise filters to
prune the combinatorial space of the ILP, 
which ensures that it can be solved in an online system.
Experiments with synthetic and real-life data demonstrate the viability of our method.

\newpage
This paper makes the following novel contributions:
\squishlist
 \item To the best of our knowledge, we are the first to
 formalize the problem of individual equity-of-attention fairness in rankings, and
define measures that capture the discrepancy between the deserved and received attention.
 \item We propose online mechanisms for fairly amortizing attention over time in consecutive rankings.
 \item We investigate the properties and behavior of the proposed mechanisms in experiments with synthetic and real-world data.
\squishend

\end{section}

\begin{section}{Equity-of-Attention Fairness}

We now formally define equity of attention accounting for {\em
  position bias}, which determines how attention is distributed over
the ranking positions. 
We consider a sequence of rankings 
at different time points, by different criteria or on request of different users.

\subsection{Notation} We use the following notation:
\begin{itemize}
\item {$u_1, ..., u_n$ is a set of subjects ranked in a system},
\item {$\rho^1, ..., \rho^m$ is a sequence of rankings},
\item {$r^j_i$ is the $[0..1]$-normalized relevance score of subject 
$u_i$ in ranking $\rho^j$},
\item {$a^j_i$ is the $[0..1]$-normalized attention value received by subject 
$u_i$ in ranking $\rho^j$},
\item $A$ denotes the distribution of cumulated attention across subjects, 
that is, $A_i = 
\sum_{j=1}^m a^j_i$ for subject $u_i$,
\item $R$ denotes the distribution of cumulated relevance across subjects, that is, $R_i = \sum_{j=1}^m r^j_i$ for subject $u_i$.
\end{itemize}

\subsection{Defining Equity of Attention}

Our fairness notion in this work is in the spirit of the {\it
  individual fairness} proposed by Dwork et. al.~\cite{dwork2012fairness}, which
requires that ``similar individuals are treated similarly'', where
``similarity'' between individuals is 
a metric capturing suitability for the task at hand.  
In the context of rankings, we consider \emph{relevance} to be a measure of subject suitability.
Further, in applications where rankings influence people's economic livelihood,
we can think of rankings not as an end, but as a means of achieving distributive justice,
that is, fair sharing of certain real-world resources.
In the context of rankings, we consider the \emph{attention of searchers} to be a resource to be distributed fairly.

There exist different types of distributive norms, one of them being \emph{equity}.
Equity encodes the idea of proportionality of inputs and outputs \cite{walster1973new},
and might be employed to account for "differences in effort, in productivity, or in contribution" \cite{yaari1984dividing}.

Building upon these ideas, we make a translational normative claim
and propose a new notion of individual fairness for rankings called
{\it equity of attention}, which requires that \emph{ranked subjects
receive attention that is proportional to their worthiness in a given search task}.
As a proxy for worthiness, we turn to the currently best available ground truth,
that is, the system-predicted relevance.


\begin{defn}[Equity of Attention]
A ranking offers equity of attention if each subject receives attention proportional to its relevance:
$$
  \frac{a_{i1}}{r_{i1}} = \frac{a_{i2}}{r_{i2}},\ \forall u_{i1}, u_{i2}.
$$
\end{defn}

Note that this definition is unlikely to be satisfied in any
single ranking, since the relevance scores of subjects are determined by
the data and the query, while the attention paid to the subjects (in terms of views or
clicks) is strongly influenced by position bias.
The effects of this mismatch will be aggravated if multiple subjects
are similarly relevant, yet obviously cannot occupy the same ranking
position and receive similar attention.

To operationalize our definition in practice, we propose an
alternative fairness definition that requires \emph{attention to be
  distributed proportionally to relevance, when amortized over a
  sequence of rankings}.

\begin{defn}[Equity of Amortized Attention]
\label{def:amortized_fairness}
A sequence of rankings $\rho^1, ..., \rho^m$ offers equity of amortized attention if each subject receives cumulative attention proportional to her cumulative relevance, i.e.:
$$
  \frac{\sum_{l=1}^{m} a^l_{i1}}{\sum_{l=1}^{m} r^l_{i1}} 
= \frac{\sum_{l=1}^{m} a^l_{i2}} {\sum_{l=1}^{m} r^l_{i2}},\ \forall u_{i1}, u_{i2}.
$$
\end{defn}

Observe that this modified fairness definition allows us to permute
individual rankings so as to satisfy fairness requirements
over time. The deficiency in the attention received by a subject
relative to her relevance in a given ranking instance can be compensated
in a subsequent ranking, where the subject is positioned higher 
relative to her relevance.

\subsection{Equality of attention}

In certain scenarios, it may be desirable for subjects to
receive the same amount of attention, irrespective of their
relevance. Such is the case 
when we suspect the ranking is biased and cannot confidently correct for that bias,
or when the subjects are not shown as an answer to any query but need to be visually displayed in a ranked order
(e.g., a list of candidates on an informational website for an election).
In such scenarios, the desired notion of fairness would be
{\it equality of attention}. 
We observe that this egalitarian version of
fairness is a special case of equity of attention, where
the relevance distributions are uniform, i.e., $r_{i1} = r_{i2} ~
\forall u_{i1}, u_{i2}$. 
As equity of attention subsumes equality
of attention, we do not explicitly discuss it further in this paper.

\subsection{Relation to group fairness in rankings}

To our knowledge, all prior works on fairness in rankings have
focused on notions of {\it group fairness}, which define fairness
requirements over the collective treatment received by all
members of a demographic group like women or men. 
Our motivation for tackling fairness at the individual level stems from the fact that
position bias affects all individuals, independently of their group
membership. It is easy to see, however, that when equity of attention
is achieved for individuals, it will also be achieved at the group level:
the cumulated attention received by all members of a group will be
proportional to their cumulated relevance. 

Prior works on fairness in rankings \cite{celis2017ranking, yang2016measuring, zehlike2017fair}
has mostly focused on diversification of the results. 
These approaches are geared for one-time rankings,
and, as any static model, will steadily accumulate equity-of-attention unfairness over time.
Since they were developed with a different goal in mind, they are not directly comparable to our dynamic approach.

Parallel with our work, Singh and Joachims
have explored similar ideas of how position bias influences fairness of exposure \cite{SinghJoachims2017} .
Their probabilistic formulations are possibly a counterpart 
of our amortization ideas, and it will be interesting to see to what extent these formulations are interchangeable.
In line with other prior works on fairness in rankings and different from our work, however, they focus
on satisfying constraints on group rather than individual fairness,
and on notions of equality rather than equity.
\end{section}

\begin{section}{Rankings With Equity of Attention}

\subsection{Measuring (un)fairness}
To be able to optimize ranking fairness, we need to measure to what extent a sequence of rankings $\rho^1, ..., \rho^m$ violates Definition~\ref{def:amortized_fairness}.
Since the proposed fairness criterion is equivalent to the requirement that the empirical distributions $A$ and $R$ 
be equal, we can measure unfairness as the distance between these two distributions. A variety of measures can be applied here, including KL-divergence, or L1-norm distance.
In this paper, measure fairness using the latter:

\begin{equation}
 \mathit{unfairness}(\rho^1, ..., \rho^m) = \sum_{i=1}^n \left| A_i - R_i \right| = \sum_{i=1}^n \left| \sum_{j=1}^m a^j_i - \sum_{j=1}^m r^j_i \right|.
 \label{eq:unfairness}
\end{equation}

L1-norm is minimized with a value of $0$ for distributions satisfying the fairness criterion from Definition~\ref{def:amortized_fairness}, 
and is thus useful as an optimization objective. 
However, since the measure is cumulative and indifferent to the exact distribution of unfairness among individuals, 
other measures could be developed to \emph{quantify} unfairness in the system at any given point. 

\subsection{Measuring ranking quality}
Permuting a ranking to satisfy fairness criteria can lead to a quality loss when 
less relevant subjects get ranked higher than more relevant ones.
We propose to quantify ranking quality using measures that draw from IR evaluation.
Traditionally, ranking models are evaluated in comparison 
with ground-truth rankings based on human-given relevance labels. 
Here, we are interested in quantifying the divergence from the original ranking.
Thus, we consider the {\em original ranking $\rho$} to be the ground-truth reference
for evaluating the quality of a {\em reordered ranking $\rho^{*}$}. 
We assume that the ground truth scores are the relevance scores
returned by the system, and that these scores reflect the best ordering of subjects. 
These considerations lead to the following definitions.

Discounted cumulative gain (DCG) quantifies the quality of a ranking by summing the relevance scores in consecutive positions
with a logarithmic discount for the values at lower positions. The measure thus puts an emphasis on having higher relevance scores at top positions.
$$
  \mathit{DCG@k}(r) = \sum_{i=1}^{k} \frac{2^{r(i)} - 1}{log_2(i+1)}
$$
This value can be further normalized by the DCG score of a perfect ranking ordered by the ground truth relevance scores.
The normalized discounted cumulative gain (NDCG)-based quality measure can be thus expressed as:
\begin{equation}
 \mathit{NDCG\mbox{-}quality}@k
 (\rho, \rho^{*}) = \frac{DCG@k(\rho^*)}{DCG@k(\rho)}
\end{equation}

This measure is maximized with a value of $1$ if the rankings do not differ or if swaps are only made within ties (i.e., subjects
with equal relevance). Other measures, like Kendall's Tau or appropriately defined $MAP\mbox{-}quality$, could be applied as well.

\subsection{Optimizing fairness-quality tradeoffs}
As discussed in the previous
section, there is ``no free lunch'': to improve fairness, we
need to perturb relevance-based rankings, which
might lead to lower ranking quality.  
To address the tradeoff, we can formulate two types
of constrained optimization problems: one where we minimize unfairness
subject to constraints on quality (i.e.,
lower-bound the minimum acceptable quality), and another where we maximize
quality subject to constraints on unfairness
(i.e., upper-bound the maximum acceptable unfairness measure).
In this paper, we focus on the former, 
since at the moment ranking quality measures are more interpretable,
and so are the constraints on quality.

\subsubsection{Offline optimization}
Let $\rho^1, ..., \rho^m$ be a sequence of rankings where the subjects are ordered by 
the relevance scores. These rankings induce zero quality loss. We wish to 
reorder them into $\rho^{1*}, ..., \rho^{m*}$ 
so as to minimize the distance between the distributions $A$ and $R$ with 
constraints on NDCG-quality loss in each ranking:

\begin{equation*}
\begin{aligned}
& \underset{}{\text{minimize}}
& & \sum_i \lvert A_i - R_i \rvert\\
& \text{subject to}
& & \mathit{NDCG\mbox{-}quality@k}(\rho^j, \rho^{j*}
) \geq \theta, \; j = 1, \ldots, m.
\end{aligned}
\end{equation*}

\noindent where $A_i$ and $R_i$ denote the cumulated attention and
relevance scores that subject $u_i$ has gained across all the $m$ rankings.

Instead of thresholding the loss in each individual ranking, an alternative would be to threshold the average loss over $m$ rankings.

\subsubsection{Online optimization}
\label{sec:online_optimization}
In practice, ranking amortization needs to be done in an online manner, one 
query at a time. Without the knowledge of future query loads, the goal is then 
to reorder the current ranking so as to minimize unfairness over 
the cumulative attention and relevance distributions in rankings seen so far, 
subject to a constraint on 
the quality of the current ranking. Thus, in the $l$-th ranking we want to
:

\begin{equation*}
\begin{aligned}
& \underset{}{\text{minimize}}
& & \sum_i \lvert A_i^{l-1} + a_i^l  ~-~  (R_i^{l-1} + r_i^l )\rvert \\
& \text{subject to}
& & \mathit{NDCG\mbox{-}quality@k}(\rho^l, \rho^{l*}) \geq \theta
\end{aligned}
\end{equation*}

\noindent where $A_i^{l-1}$ and $R_i^{l-1}$ denote the cumulated
attention and relevance scores that subject $u_i$ has gained up to and
including ranking $\rho^{l-1}$.

\subsection{An ILP-based fair ranking mechanism}
\label{sec:ilp}
\subsubsection{ILP for online attention amortization}
The optimization problem defined in Sec.~\ref{sec:online_optimization} can be solved as an integer linear program (ILP).
Assume we are to rerank the $l$-th ranking in a series of rankings.
We introduce $n^2$ decision variables $X_{i,j}$ which are set to $1$ if subject $u_i$ is assigned to the ranking position $j$,
and set to 0 otherwise.
At the time of reordering the $l$-th ranking, the following values are constants:
\begin{itemize}
\item relevance scores for each subject $u_i$ in the current ranking: $r_i^l$,
\item attention values assigned to ranking positions: $w_j$,
\item relevance scores accumulated up to (and excluding) the current ranking for each subject: $R_i^{l-1}$,
\item attention values accumulated up to (and excluding) the current ranking for each subject: $A_i^{l-1}$,
\item IDCG@k value computed over the current ranking $\rho_l$, which is used as a normalization score for NDCG-quality@k.
\end{itemize}

For each subject $u_i$, the accumulated attention and relevance are
initialized as $A_i^0 = 0$ and $R_i^0 = 0$ for all $u_i$.

The ILP is then defined as follows:
\begin{equation}
\begin{aligned}
& \underset{}{\text{minimize}}
& & \sum_{i=1}^n \sum_{j=1}^n \lvert A_i^{l-1} + w_j - (R_i^{l-1} + r_i^l) \rvert \cdot X_{i,j} \\
& \text{subject to}
& & \sum_{j=1}^k \sum_{i=1}^n {2^{r_i^{l}} - 1 \over log_2(j+1) } X_{i,j} \geq \theta \cdot IDCG@k \\
& & & X_{i,j} \in \{0,1\}, \; \forall_{i,j} \\
& & & \sum_i X_{i,j} = 1, \; \forall_j \\
& & & \sum_j X_{i,j} = 1, \; \forall_i
\end{aligned}
\label{eq:ilp}
\end{equation}
The first constraint bounds the loss in ranking quality, in terms of the NDCG-quality measure, by the multiplicative threshold $0 \leq \theta \leq 1$.
The other constraints ensure that the solution is a bijective mapping of subjects onto ranking positions.
The terms $A_i^{l-1} + w_j$ and $R_i^{l-1} + r_i^l$ encode the updates of the cumulative attention and relevance, respectively,
if and only if $u_i$ is mapped to position $j$.

It is worth noting that:
\begin{itemize}
\item {When $\theta = 1$, we do not allow any quality loss.
This, however, does not mean that the ranking will remain unchanged.
Subjects can be reordered within ties to minimize unfairness.}
\item When $\theta = 0$, any permutation of the ranking is allowed
striving to minimize unfairness in the current iteration.
\end{itemize}

\subsubsection{ILP with candidate pre-filtering}
\label{sec:ilp-filtering}
The ILP operates on a huge combinatorial space, with the number of binary variables
being quadratic in the number of subjects.
Real systems deal with millions of subjects, and the optimization needs to be
carried out each time a new ranking is requested.
Such a problem size is a bottleneck for ILP solvers,
and in practice the optimization needs
to use approximation algorithms, such as
LP relaxations or greedy-style heuristics. 
This is one of the directions for further research.

To deal with the issue in this paper, 
instead of reranking all subjects in each iteration, we rerank only subjects from a prefiltered candidate set. 
Different strategies are possible for selecting the candidate sets. 
On the one hand, prefiltering the top-ranked subjects by relevance scores would let us satisfy the quality constraints, but 
may entail small fairness gains, especially for near-uniform relevance distributions.
On the other hand, prefiltering based on the objective function might lead to situations where the ILP cannot 
find any solution without violating the constraints.
\footnote{Without prefiltering, the ILP 
always has at least one feasible solution (the original ranking).}

Our strategy thus is as follows. Assume we want to select a subject candidate subset $D$ of size $t$ to be reranked, 
and we constrain the quality in Eq.~\ref{eq:ilp} at rank $k$. Since the attention weights $w_j$ are positive, 
the biggest contributors to the objective function are the subjects with the smallest values of
$A_i - (R_i + r_i)$. These are the subjects with the highest deficit (negative value) of fair-share attention.
We always select $k$ subjects with the highest relevance scores in $r^l$,
to make sure we satisfy the quality constraint,
plus other $t-k$ subjects with the lowest $A_i - (R_i + r_i)$ values, who
are most worthy of being promoted to high ranks.
As a result, when no feasible solution can be found by reranking the most worthy subjects, 
the ILP will default to choosing the top-$k$ candidates by relevance scores.

\subsubsection{Extensions}

\paragraph{Granularity.}
The presented model assumes
that attention and relevance are aggregated per ranked subject. It is
straightforward to extend it to handle higher-level actors
such as product brands or Internet domains, by summing the relevance and attention scores over the corresponding subjects. 
As a consequence of this modification, bigger organizations would obtain higher exposure. Deciding whether this
effect is fair is a policy issue. 

\paragraph{Handling dynamics.}
In a real-world system, the size of the population will vary over time,
with new subjects joining and existing ones dropping out.
Our model is capable of handling this kind of dynamics,
since new users starting with no deserved attention will be positioned in between the users who got more than they deserved and those who got less. 
Moreover, ranking quality constraints will prevent such users from being positioned too low in rankings where they are highly relevant.
 
\end{section}

\section{Experiments}

\subsection{Data}
The datasets we use are either synthetically generated or derived from other publicly available resources.
They are freely available to other researchers.

\subsubsection{Synthetic datasets.}
We create 3 synthetic datasets to analyze the performance of the model in a controlled setup
under  different relevance distributions. We assume the following distribution shapes: (i) \textbf{uniform}, where every user has the same relevance score,
(ii) \textbf{linear}, where the scores decrease linearly with the rank position, and (iii) \textbf{exponential}, where the scores decrease exponentially with the rank position.
Each dataset has $100$ subjects.

\subsubsection{Airbnb datasets.}
To analyze the model in a real-world scenario,
we construct rankings based on Airbnb\footnote{\url{https://www.airbnb.com/}} apartment listings from 3 cities located in different parts of the world:
Boston, Geneva, and Hong Kong. Airbnb is a two-sided sharing economy platform allowing people to offer their free rooms or apartments for short-term rental.
It is a prime example of a platform where exposure and attention play a crucial role in 
the subjects' financial success.
The data we use is freely available for research.\footnote{Downloaded from \url{http://insideairbnb.com/}}

Rankings are constructed using the attribute $id$ as a subject identifier, 
and various review ratings as the ranking criteria, with the rating scores serving as relevance scores.
Such crowd-sourced judgments serve as a good worthiness-of-attention proxy on this particular platform,
although one has to have in mind that rating distributions tend to be skewed towards higher scores,
which is confirmed by our experimental analysis.

For each of the $3$ datasets, we run the amortization model on two types of ranking sequences:
\begin{enumerate}
 \item {\textbf{Single-query}: We examine the amortization effects when a single ranking is repeated multiple times. 
 To construct the rankings, we use the values of the $review\_scores\_rating$ attribute, which corresponds to the overall quality of the listing.}
 \item {\textbf{Multi-query}: We examine the behavior of the model when a sequence of rankings, each with a different relevance distribution, is repeated multiple times.
 To this end, for each city, we construct 7 rankings based on different rating attributes: $review\_scores\_rating$,\\
 $review\_scores\_accuracy$, $review\_scores\_cleanliness$,\\ 
 $review\_scores\_checkin$, $review\_scores\_communication$,\\ 
  $review\_scores\_location$, 
  and $review\_scores\_value$.}
\end{enumerate}

The datasets for Boston, Geneva, and Hong Kong contain $3944$, $1728$, and $4529$ subjects, respectively.

Note that, for the purpose of model performance evaluation, the queries themselves become irrelevant once the relevance is computed.
Since the values of the aforementioned attributes serve as relevance scores, the queries are abstracted out.

\subsubsection{StackExchange dataset.}
We create another dataset from a querylog and a document collection synthesized from the StackExchange dump by Biega et al. \cite{biega2017privacy},
please refer to the original paper for details. 
We choose a radom subset of users and order their queries by timestamps, creating a workload of around 20K queries.
We use Indri\footnote{\url{https://www.lemurproject.org/indri/}} to retrieve 500 most relevant answers for each query, and treat the author of the answer as the subject to be ranked.
Using this dataset helps us gain an insight into the performance of the method in core IR tasks and with different sets of subjects ranked in each iteration.

\subsection{Position bias}
Our model requires that we assign a weight to each ranking position, denoting the fraction of the total attention the position gets.
These weights will depend on the application and platform, and may be estimated from historical click data.
In this paper we study the behavior of the 
equity-of-attention mechanism 
under generic models of attention distribution.
We focus on the following distributions:
\begin{enumerate}
 \item {\textbf{Geometric}: The weights of the positions are distributed geometrically with the parameter $p$ up to the position $k$, and are $0$ for positions lower than $k$.
 Geometrically distributed weights are a special case of the cascade model \cite{craswell2008experimental}, where each subject has the same probability $p$ of being clicked.
 Setting the weights of lower positions to $0$ is based on an assumption that low-ranked subjects are not inspected.
\begin{equation}
w_j = \begin{cases} 
      p(1-p)^{j-1} & j \leq k \\
      0 & j > k \\
   \end{cases}
\label{eq:attention_geometric}
\end{equation}}
 \item {\textbf{Singular}: The top-ranked subject receives all the attention. This is a special case of the geometric attention model with parameters $p=1, k=1$. Studying this attention model is motivated by systems such as Uber, 
 which present only top-1 matches to the searchers by default.
\begin{equation}
w_j = \begin{cases} 
      1 & j = 1 \\
      0 & j > 1 \\
   \end{cases}
\end{equation}}

\end{enumerate}
Before being passed on to the model, the weights are rescaled such that $\sum_{j} w_j = 1$.
Studying the effects of position bias on individual fairness under more complex attention models is future work.

\subsection{Implementation and parameters}
We implement the ILP-based amortization defined in Section~\ref{sec:ilp} using the Gurobi software.\footnote{\url{http://www.gurobi.com/}}
Constraints are set to be satisfied up to a feasibility threshold of $1e-7$.
We prefilter 100 candidates for reranking in each iteration, as described in Section~\ref{sec:ilp-filtering}.

In the singular attention model, since all the attention is assumed to go to the first ranking position, the ILP constrains the NDCG-quality at rank $k=1$.
We construct the geometric attention model with $p=0.5$ and $k=5$, and in this case the
ILP constraints the NDCG-quality at rank $k=5$.

In the single-query mode, where a single ranking is repeated multiple times, we set the number of iterations to $20K$.
In the multi-query mode, with a repeated sequence of different rankings, we repeat the whole sequence $3K$ times,
which leads to a total of $21K$ rankings.

Relevance scores in the framework need to be normalized to form a distribution.
In this paper, we assume relevance is a direct proxy for worthiness and rescale the rating scores linearly.
Note, however, that if additional knowledge is available to the platform regarding the correspondence
between relevance and worthiness, other transformations can be applied as well.

\subsection{Mechanisms under comparison}
We compare the performance of the ILP-based online mechanism
against two baseline heuristics.
\begin{enumerate}
 \item {\textbf{Relevance}: The first heuristic is to allow only relevance-based ranking, completely disregarding fairness.}
 \item {\textbf{Objective}: The second heuristics is an objective-driven ranking
strategy, which orders subjects by the increasing priority value: $A_i - R_i - r_i$ (see Sec.~\ref{sec:ilp-filtering})
for each ranking.
Since all position weights $w_j$ are positive, assigning highest weights to 
subjects with the lowest preference value is in line with the minimization goal.
This ranking strategy aims at strong fairness amortization without any quality constraints, 
and is expected to perform similarly to the ILP with $\theta=0$.}
\end{enumerate}

\subsection{Data characteristics: relevance vs. attention}
Figure~\ref{fig:relevance_dists} shows the relevance score distributions
in the single-query Airbnb datasets for Boston, Geneva, and Hong Kong. 
The seemingly flatter shape of the Boston and Hong Kong distributions
is the result of a bigger size of these datasets when compared to the Geneva dataset, where each individual has, on average, a larger fraction of the total relevance.
Overall, the distributions have a shape which complements the uniform, linear, and exponential shapes of distributions in the synthetic datasets.

Figure~\ref{fig:relevance_vs_attention} presents an example strongly motivating our research.
Namely, it compares the distribution of relevance in the Geneva dataset with the distribution
of attention according to the geometric model with $p=0.5$, where the weights closely follow
the empirical observations made in previous position bias studies \cite{JoachimsRadlinski2007}.
Observe that the relevance distribution plotted in green is the same as that in Figure~\ref{fig:relevance_dists}.
There is a huge discrepancy between these two distributions, while, as argued in this paper, they should ideally be equal to ensure individual fairness.
Similar discrepancy exists in the two other Airbnb datasets.

\begin{figure}[bp!]
	\centering
	\includegraphics[width=0.65\linewidth]{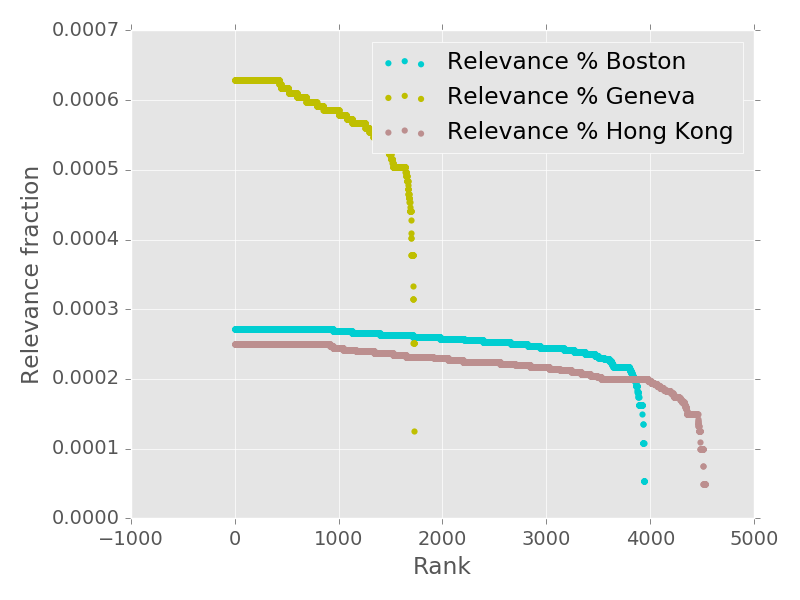}
	\caption{Relevance distributions in the Airbnb datasets.}
	\label{fig:relevance_dists}
\end{figure}

\begin{figure}[bp!]
	\centering
	\includegraphics[width=0.65\linewidth]{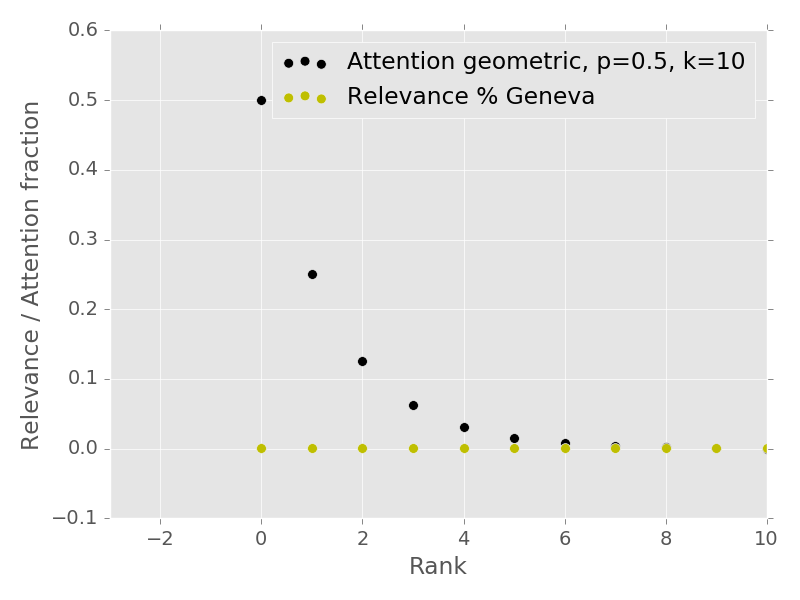}
	\caption{Comparison of the attention and relevance distributions for the top-10 ranking positions in the Geneva dataset.
	Note that the relevance distribution presented here is the same as in Fig.~\ref{fig:relevance_dists}. 
	To satisfy equity-of-attention fairness, the two distributions would have to be the same.}
	\label{fig:relevance_vs_attention}
\end{figure}

\subsection{Performance on synthetic data}
\paragraph{Singular attention model}
Figure~\ref{fig:uni-singular} reveals a number of interesting properties of the mechanism for the Uniform relevance distribution.
We plot the iteration number on the x-axis, and the value of the unfairness measure defined by Equation~\ref{eq:unfairness} on the y-axis.
First, since reshuffling does not lead to any quality loss when all the relevance scores are equal, all the reshuffling methods perform equally well irrespective of $\theta$.
Their amortizing behavior should be contrasted with the black line denoting the relevance baseline. 
Unfairness for this method always increases linearly by a constant factor incurred by the single ranking.
Second, amortization methods periodically bring unfairness to 0. The minimum occurs every $n$ iterations, where $n$ is the number of subjects in the dataset.
Within the cycle, each subject is placed in the top position (receiving all the attention) exactly once.

Figure~\ref{fig:lin-singular} with the results for the Linear dataset,
confirms another anticipated behavior. With no ties in the relevance scores, it is not possible to improve fairness without incurring quality loss.
Thus, all methods with $\theta > 0$ lead to higher unfairness when compared to the Objective baseline,
although the unfairness is still lower in ILP with $\theta<0.8$ than in the Relevance baseline.

When the relevance scores decrease exponentially (Figure~\ref{fig:exp-singular}), the ILP is not able to satisfy the quality constraint with any $\theta>=0.5$,
and thus these rerankings become equivalent to those of the Relevance heuristic.

\begin{figure*}[th!]
    \centering
    \begin{minipage}{0.3\textwidth}
        \centering
        \includegraphics[width=1.0\textwidth]{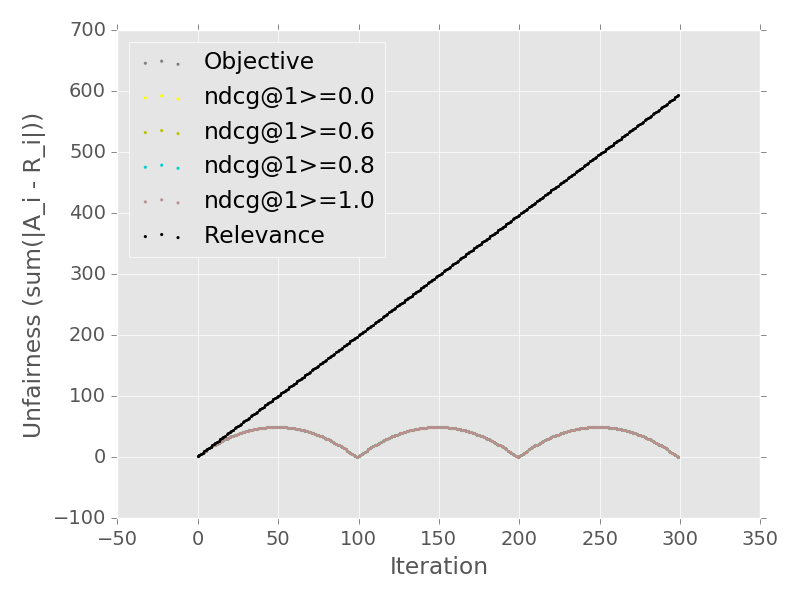}
        \vspace{-20px}
        \caption{Model performance on the synthetic Uniform dataset. Attention singular.}
        \label{fig:uni-singular}
    \end{minipage}\hfill
    \begin{minipage}{0.3\textwidth}
        \centering
        \includegraphics[width=1.0\textwidth]{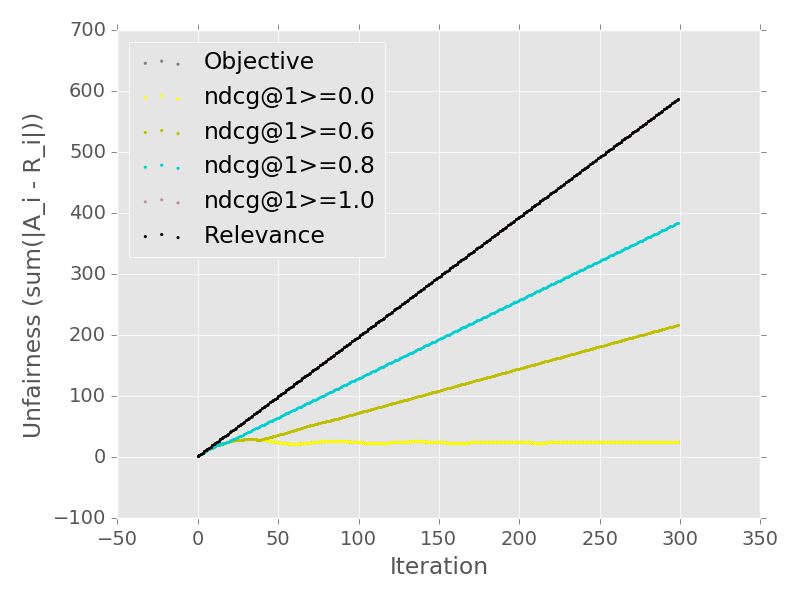}
        \vspace{-20px}
        \caption{Model performance on the synthetic Linear dataset. Attention singular.}
        \label{fig:lin-singular}
    \end{minipage}\hfill
    \begin{minipage}{0.3\textwidth}
        \centering
        \includegraphics[width=1.0\textwidth]{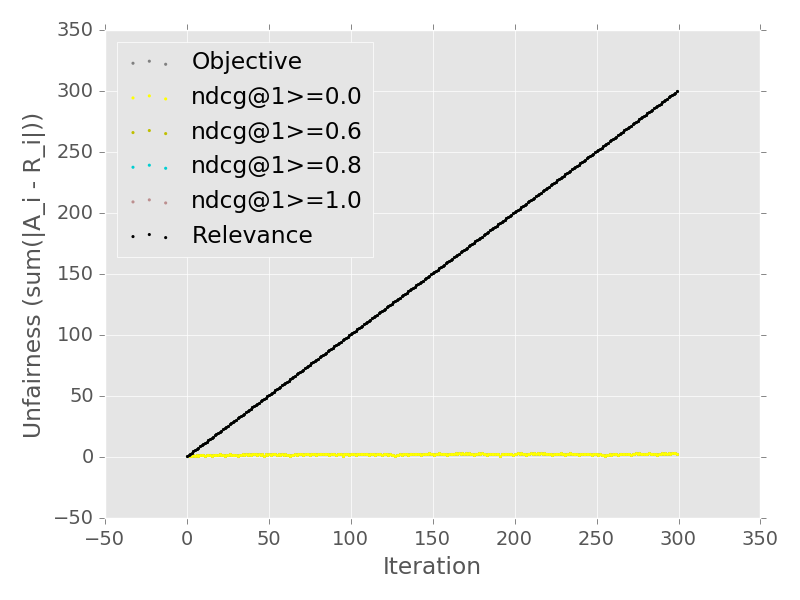}
        \vspace{-20px}
        \caption{Model performance on the synthetic Exponential dataset. Attention singular.}
        \label{fig:exp-singular}
    \end{minipage}
    
    \begin{minipage}{0.3\textwidth}
        \centering
        \includegraphics[width=1.0\textwidth]{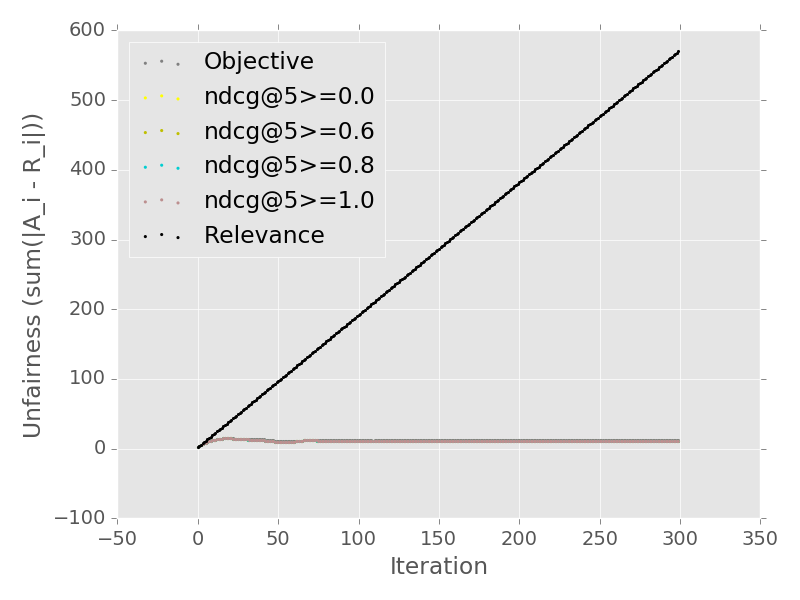}
        \vspace{-20px}
        \caption{Model performance on the synthetic Uniform dataset. Attention geometric.}
        \label{fig:uni-geometric}
    \end{minipage}\hfill
    \begin{minipage}{0.3\textwidth}
        \centering
        \includegraphics[width=1.0\textwidth]{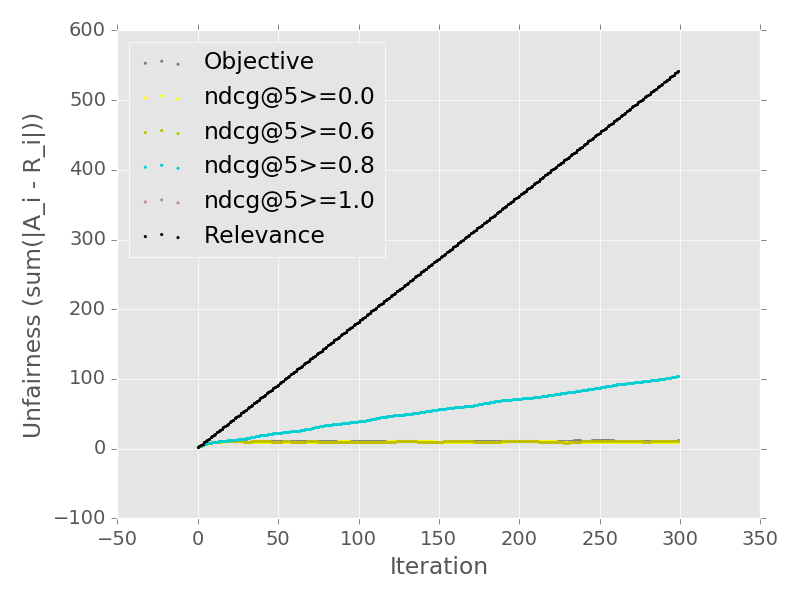}
        \vspace{-20px}
        \caption{Model performance on the synthetic Linear dataset. Attention geometric.}
        \label{fig:lin-geometric}
    \end{minipage}\hfill
    \begin{minipage}{0.3\textwidth}
        \centering
        \includegraphics[width=1.0\textwidth]{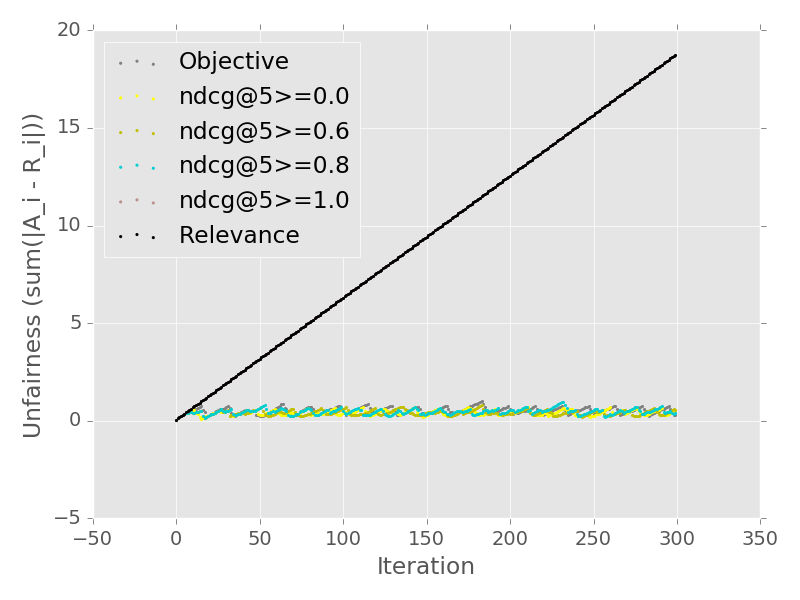}
        \vspace{-20px}
        \caption{Model performance on the synthetic Exponential dataset. Attention geometric.}
        \label{fig:exp-geometric}
    \end{minipage}
\end{figure*}

\vspace{-5px}
\paragraph{Geometric attention model}
As shown in Figures~\ref{fig:uni-geometric}, \ref{fig:lin-geometric}, and \ref{fig:exp-geometric},
the periodicity effect becomes less pronounced under the general geometric attention model.
Figure~\ref{fig:equal_geom_kparam} helps to understand this behavior
by showing the unfairness values achieved by the Objective heuristic with different values of the attention cut-off $k$
(see Equation~\ref{eq:attention_geometric}).
With $k=1$, the model is equivalent to Singular. As we increase $k$, the distribution of the position weights becomes smoother,
smoothing also the periodicity of the unfairness values.

The very good performance of the ILP-based rerankings with any $\theta<1$ in Figure~\ref{fig:exp-geometric},
stems from the fact that the relevance and attention distributions are almost the same 
(the only difference being that the scores in the relevance distribution are non-zero for more positions).
Our results show that in this case the ILP performs a reordering only every now and then, when the subjects ranked lower than position 5 in the original ranking
gather enough deserved attention. This causes the unfairness to go up and down periodically.

\begin{figure}[bp!]
\vspace{-15px}
	\centering
	\includegraphics[width=0.65\linewidth]{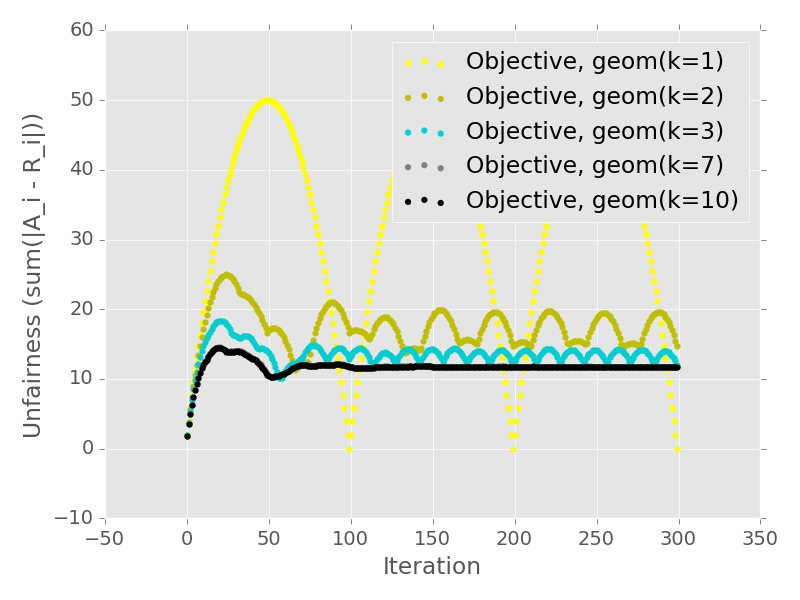}
	\caption{Performance of the Objective heuristic on the synthetic Uniform dataset under the geometric attention model with different attention cut-off points.}
	\label{fig:equal_geom_kparam}
\end{figure}

\subsection{Performance on Airbnb data}
\subsubsection{Single-query, singular attention.}
We first analyze the model performance on the Airbnb datasets where a single ranking is repeated multiple times,
and the attention model is set to singular.
The results are shown in Figures~\ref{fig:boston-single-singular}, \ref{fig:geneva-single-singular}, \ref{fig:hongkong-single-singular}
for Boston, Geneva, and Hong Kong, respectively. 
As in the analysis with the synthetic data, we plot the iteration number on the x-axis, and the value of the unfairness measure defined by Equation~\ref{eq:unfairness} on the y-axis.
There are a number of observations:
\begin{itemize}
 \item {As noted before, the loss in the Relevance baseline (plotted in black) increases linearly by the constant unfairness factor incurred by the single ranking.}
 \item {Relaxing the quality constraint by decreasing $\theta$ allows us to achieve lower unfairness values in the corresponding ranking iterations.}
 \item {The Objective heuristic with no quality constraints and the ILP where $\theta=0$ are able to amortize fairness over time well, with no significant growth of unfairness over time.}
 \item {The periodicity effect we observed on synthetic uniform data appears here as well. This is due to the relative closeness of the relevance distributions in the Airbnb data to the uniform distribution.
 Unfairness achieved by the amortizing methods is close to $0$ every $n$ iterations. The frequency of the minimum indeed corresponds to the size of the respective datasets.}
 \item {In some methods unfairness starts to grow linearly after a certain number of iterations (see, e.g., the blue curve in Figure~\ref{fig:boston-single-singular}). 
 This is a side effect of the candidate prefiltering heuristic we chose. When the ILP receives a filtered candidate set where no subjects filtered based on the objective can be placed 
 at the top of the ranking without violating the quality constraint, the ILP defaults to placing the most relevant subjects at the top, which causes the quality loss to be $0$ and the unfairness growing linearly.
 This effect persists until some of the more relevant subjects gather enough deserved attention to be pre-selected - note the variability that occurs in the blue curve again starting around the 17K-th iteration.}
 \item {For a number of iterations at the beginning (equal to the number of ties at the top of the ranking), all the methods perform the same, irrespective of the quality constraints.  
 This is due to the fact that unfairness is minimized by reshuffling the most deserving relevant subjects first, which does not incur any quality loss.}
\end{itemize}

\begin{figure*}[th!]
    \centering
    \begin{minipage}{0.3\textwidth}
        \centering
        \includegraphics[width=1.0\textwidth]{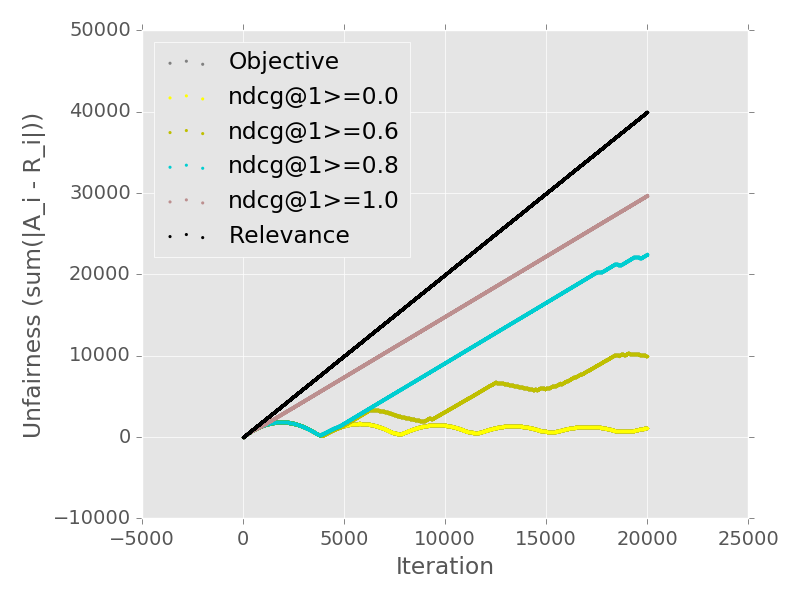}
        \vspace{-20px}
        \caption{Model performance on the single-query Boston dataset. Attention singular.}
        \label{fig:boston-single-singular}
    \end{minipage}\hfill
    \begin{minipage}{0.3\textwidth}
        \centering
        \includegraphics[width=1.0\textwidth]{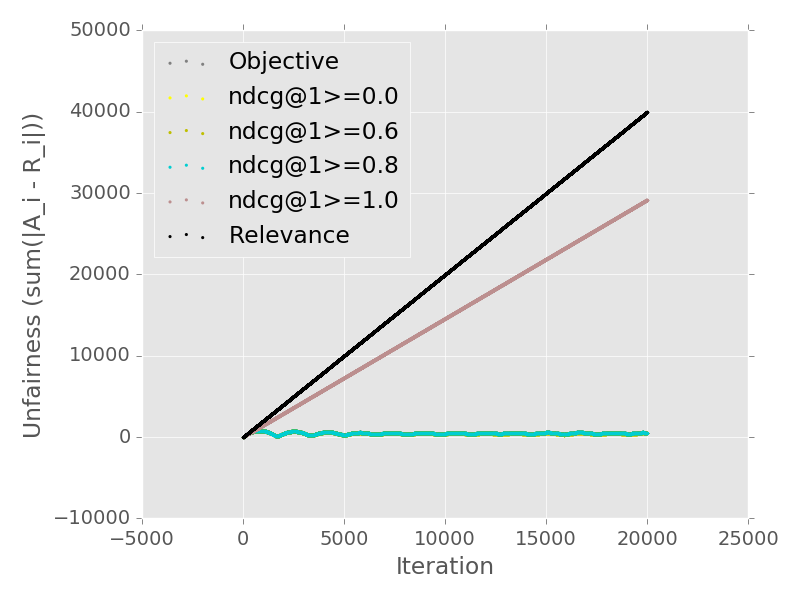}
        \vspace{-20px}
        \caption{Model performance on the single-query Geneva dataset. Attention singular.}
        \label{fig:geneva-single-singular}
    \end{minipage}\hfill
    \begin{minipage}{0.3\textwidth}
        \centering
        \includegraphics[width=1.0\textwidth]{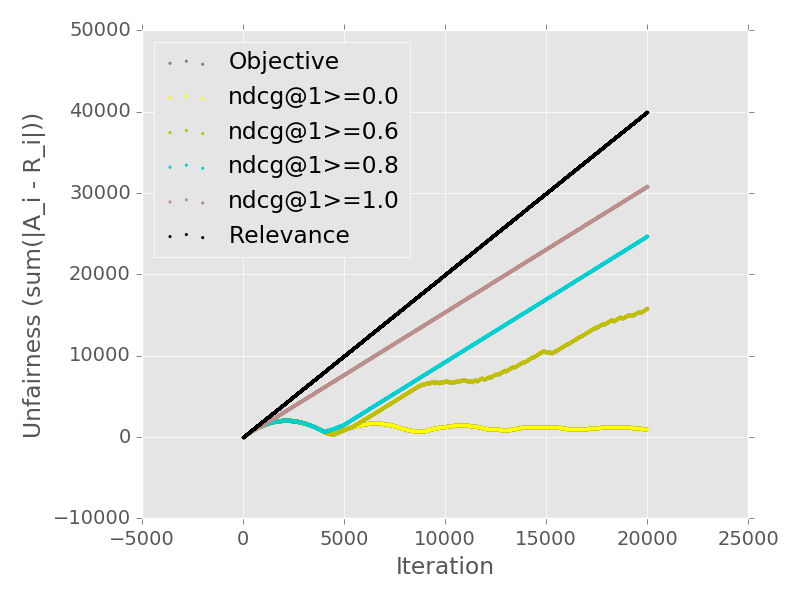}
        \vspace{-20px}
        \caption{Model performance on the single-query Hong Kong dataset. Attention singular.}
        \label{fig:hongkong-single-singular}
    \end{minipage}
    
    \begin{minipage}{0.3\textwidth}
        \centering
        \includegraphics[width=1.0\textwidth]{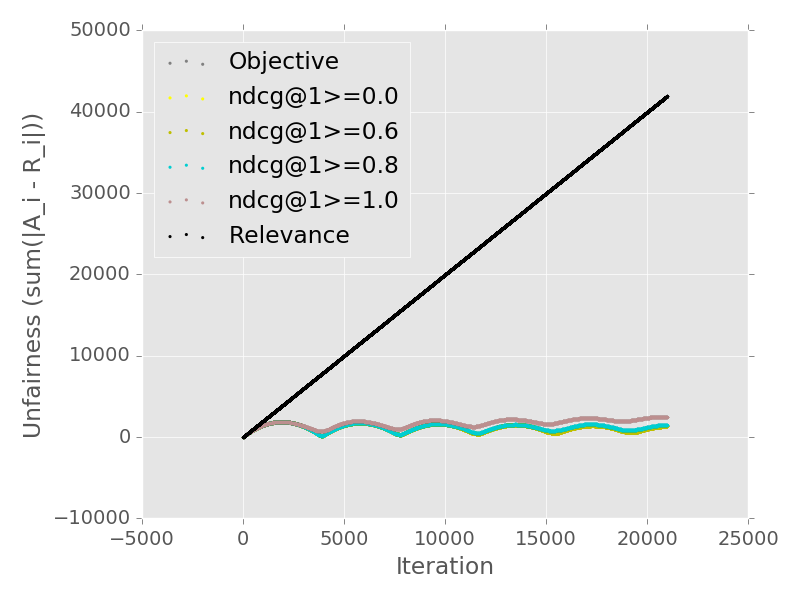}
        \vspace{-20px}
        \caption{Model performance on the multi-query Boston dataset. Attention singular.}
        \label{fig:boston-multi-singular}
    \end{minipage}\hfill
    \begin{minipage}{0.3\textwidth}
        \centering
        \includegraphics[width=1.0\textwidth]{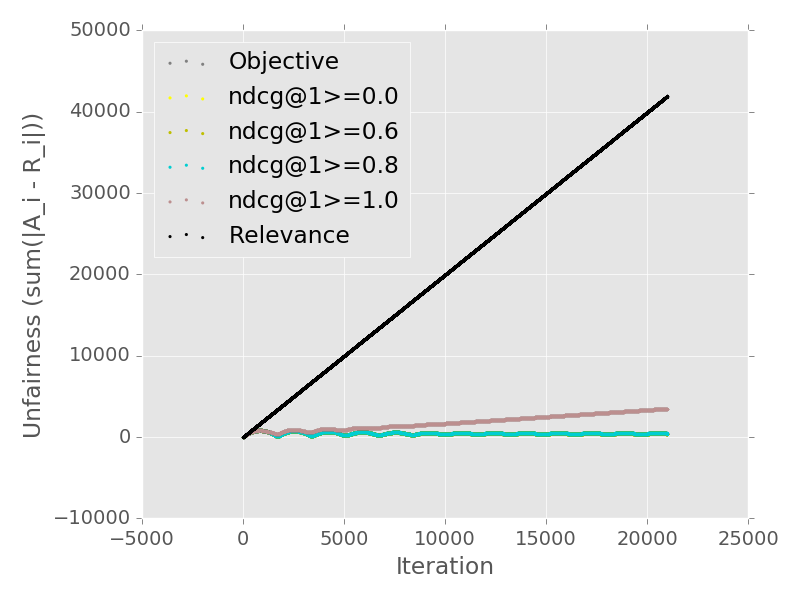}
        \vspace{-20px}
        \caption{Model performance on the multi-query Geneva dataset. Attention singular.}
        \label{fig:geneva-multi-singular}
    \end{minipage}\hfill
    \begin{minipage}{0.3\textwidth}
        \centering
        \includegraphics[width=1.0\textwidth]{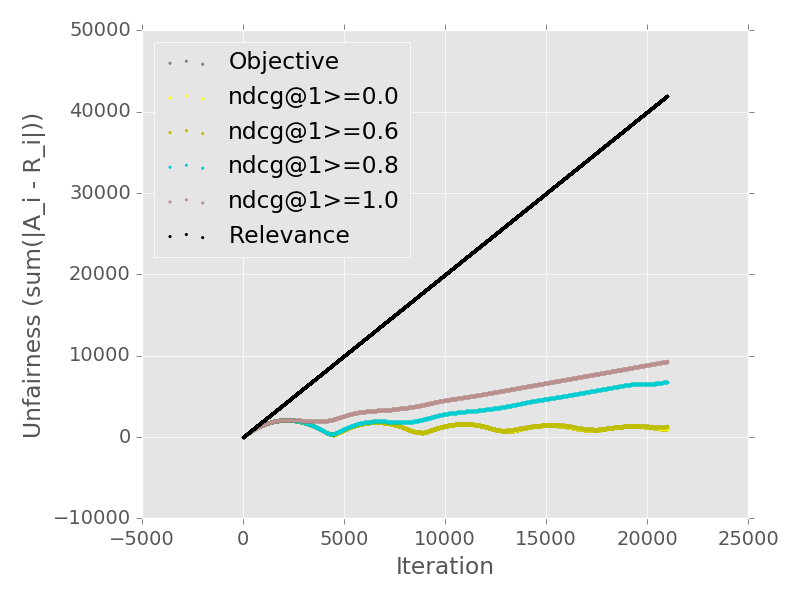}
        \vspace{-20px}
        \caption{Model performance on the multi-query Hong Kong dataset. Attention singular.}
        \label{fig:hongkong-multi-singular}
    \end{minipage}
\end{figure*}

\subsubsection{Multi-query, singular attention.}
Our methods amortize fairness better (achieving lower unfairness) on the Airbnb multi-query datasets
(Figures~\ref{fig:boston-multi-singular}, \ref{fig:geneva-multi-singular}, and \ref{fig:hongkong-multi-singular}) when compared to the single-query datasets for two reasons.
First, the variability in subject relevance and ordering in different iterations is a factor helpful in smoothing the deserved attention distributions over time.
Second, distributions of the rating attributes in the Airbnb datasets used to construct the rankings are more uniform than the global rating score,
and have more ties at the top of the ranking. 
These relevance distribution characteristics enable methods with conservative quality constraints (even  the ILP with $\theta=1$) to perform very well. 

\subsubsection{Single-query, geometric attention.}
The general geometric attention distribution is closer to the relevance distributions in the Airbnb datasets than the singular distribution is.
As noted in the analysis with synthetic data, the closeness of the two distributions helps amortize fairness at a lower quality loss.
We can observe a similar effect in Figure~\ref{fig:boston-single-geometric}, with more ILP-based methods reaching the performance of the Objective heuristic.
Note, however, that the improved performance here is also partly due to the fact that we constrain the quality at a higher rank when assuming the geometric attention,
which is easier to satisfy.

\begin{figure}[bp!]
\vspace{-3px}
	\centering
	\includegraphics[width=0.65\linewidth]{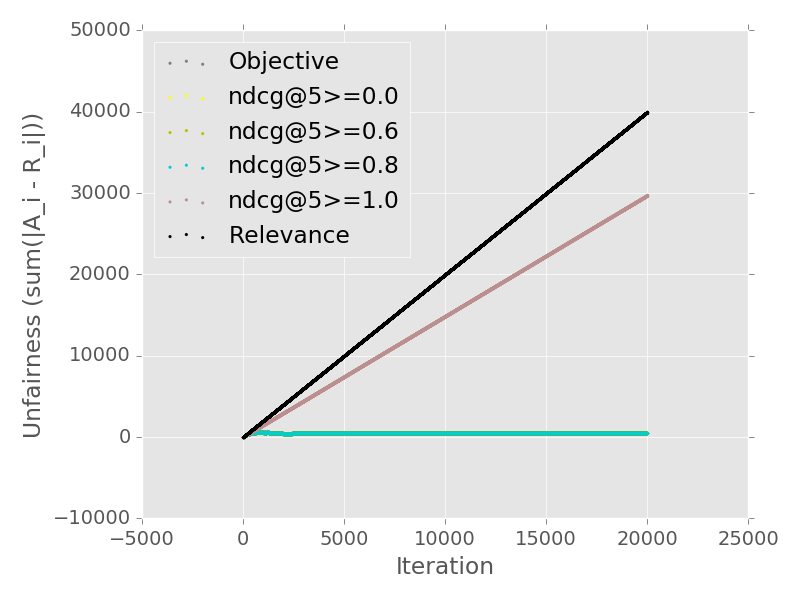}
        \caption{Model performance on the single-query Boston dataset. Attention geometric. Results are similar for the Geneva and Hong Kong datasets.}
        \label{fig:boston-single-geometric}
\end{figure}

\subsubsection{Unfairness vs. quality loss.}
The results presented so far show the performance of the ILP-based fairness amortization under different quality thresholds.
Since the thresholds bound the maximum quality loss over all iterations, the actual loss in most cases might be lower.
To investigate these effects, we plot the actual NDCG-quality values of the rerankings done by different methods on the Boston dataset under the Singular attention model in Figure~\ref{fig:quality_boston_singular}.
The results confirm that the actual loss is often lower than the threshold enforced by the ILP.
Observe that NDCG-quality is $1$ for a number of initial iterations in all the methods. This is where reshuffling of the top ties happens.
The quality starts decreasing as less relevant subjects gather enough deserved attention, and periodically goes back to 1, when the top-relevant subjects gain priority again.
Similar conclusions regarding the absolute loss hold under the general geometric attention model.

\begin{figure}[bp!]
	\centering
	\includegraphics[width=0.65\linewidth]{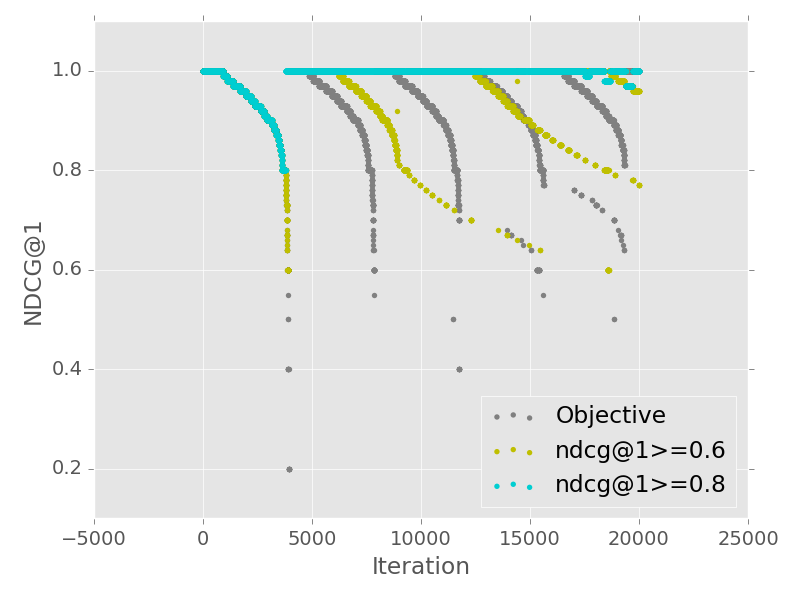}
	\caption{Actual values of ranking quality. Boston dataset, attention singular.}
	\label{fig:quality_boston_singular}
\end{figure}

\subsection{Performance on StackExchange data}
The relative trends in the performance of our method are the same here as in the results for other datasets.
One of the characteristics that distinguish the StackExchange dataset
is that each individual subject occurs in relatively few rankings. 
An observation that follows is that longer amortization timeframe is necessary under such conditions
- a subject obviously needs to appear in a number of rankings 
so that the model can reposition them to fairly distribute attention.

\begin{section}{Related work}

\noindent{\bf Fairness.}
The growing ubiquity of data-driven learning models in algorithmic
decision-making has recently boosted concerns about the issues of fairness and bias
(see, e.g., \cite{FATconference} and the pointers there).
The problem of discrimination in data mining and machine learning
has been studied for a number of years 
(e.g., \cite{DBLP:conf/kdd/PedreschiRT08,
DBLP:conf/pkdd/KamishimaAAS12,DBLP:journals/ker/RomeiR14}).
The goal there is to analyze and counter data bias and 
unfair decisions that may lead to discrimination.
Much prior work has centered around various notions of group fairness:
preserving certain ratios of members of protected vs. unprotected 
groups in the decision making outcomes,
with the groups derived from discrimination-prone attributes like gender,
race, nationality, etc.
\cite{Feldman:KDD2015,DBLP:conf/nips/HardtPNS16}.
For example, the criterion of statistical parity requires 
that a classifier's outcomes do not depend on the
membership in the protected group.
State-of-the-art mechanisms for dealing with such
group fairness requirements are
to solve constrained optimization, e.g. maximize
prediction accuracy subject to certain bounds on group membership
in the output labels. This has led to classification models with fairness-aware
regularization (e.g., \cite{DBLP:conf/www/ZafarVGG17}).
Beyond the fairness of outcomes, researchers have looked into the fairness of process in the decision-making systems \cite{grgic2018beyond}.

Individual fairness \cite{dwork2012fairness}
requires that individual subjects who have similar attributes
should, with high probability, receive the same
prediction outcomes. Literature to this
end has so far focused on classification and selection problems
\cite{DBLP:conf/icml/ZemelWSPD13,Kearns:ICML2017}. 

Other lines of work investigate mechanisms for fair division of resources \cite{abebe2017fair}, 
or how automated systems can assist humans in decision making \cite{kleinberg2017human}.

\noindent{\bf Fairness in rankings.}
Prior work on fair rankings is scarce and recent.
Some proposals show how to incorporate
various notions of group fairness into ranking quality measures
\cite{yang2016measuring}.
There have been approaches that 
diversify the ranking results in terms of presence of members of different groups in ranking prefixes,
at the same time keeping the ranking quality high
 \cite{zehlike2017fair}.
This problem has also been studied from a theoretical perspective
with the results provided for the computational complexity of the problem
\cite{celis2017ranking}.
All of these approaches consider static rankings only,
and all focus on group fairness.
Parallel with our work, Singh and Joachims \cite{SinghJoachims2017} have proposed a
notion of group fairness based on equality of exposure for demographic groups.
While technically complementary and similar in spirit to our approach,
this method is also geared for a purpose different than individual fairness,
and does not aim at binding attention to relevance.

\noindent{\bf Bias in IR.}
The existence of position bias in rankings of search results has been revealed by
a number of eye-tracking and other empirical studies \cite{craswell2008experimental,
DBLP:conf/sigir/DupretP08,DBLP:conf/wsdm/GuoLW09}.
Top-ranked answers have a much higher probability of being
viewed and clicked than those at lower ranks. 
The effect persist even if the elements at different ranks
are randomly permuted
\cite{JoachimsRadlinski2007}.
These observations have led to a variety of click models
(see \cite{chuklin2015click} for a survey), and several methods for
bias-aware re-ranking (e.g., \cite{DBLP:conf/sigir/WangBMN16,
DBLP:conf/wsdm/JoachimsSS17}).
However, position bias has been primarily studied in the context of document ranking
and no prior work has investigated the influence of the bias on the fairness of ranked results.
A large search engine has been investigated for presence of
differential quality of results across demographic groups \cite{Mehrotra:WWW2017}.
Similar studies have been carried out on other kinds of
tasks such as credit worthiness or recidivism prediction \cite{Adler:ICDM2016}.

\noindent{\bf Relation to other models.}
Fairness dimension has been considered for job dispatching at the OS level,
for packet-level network flows \cite{ghodsi2012multi},
for production planning in factories \cite{Ghodsi:NSDI2011},
and even for two-sided matchmaking in call centers \cite{Armony:OR2010}.
Fairness understood as envy-freeness is also investigated in computational advertising,
including generalized second-price auctions \cite{edelman2007internet}.
In the context of rankings, a potential connection
between fair rankings and fair queuing has recently been suggested \cite{chakraborty2017fair}.

\end{section}

\section{Conclusion}

\balance

This paper argues for equity of attention -- a new notion of fairness in rankings,
which requires that the attention ranked subjects receive from searchers
is proportional to their relevance. As this definition cannot be satisfied in 
a single ranking because of the position bias, 
we propose to amortize fairness over time
by reordering consecutive rankings, 
and formulate a constrained optimization problem which achieves this goal.

Our experimental study using real-world data shows that the discrepancy 
between the attention received from searchers and the deserved attention can be substantial,
and that many subjects have equal relevance scores.
These observations suggest that improving equity of attention is crucial
and can often be done without sacrificing much quality in the rankings.
Incorporating such fairness mechanisms is especially important 
on sharing economy or two-sided market platforms where rankings influence
people's economic livelihood, and our work addresses this gap.

Equity of attention opens a number of interesting directions for future work,
including calibration of ranker scores in economically-themed applications,
all the way down the IR stack to properly training judges to provide relevance labels with fairness in mind.

\ \\
\noindent \small { \textbf{Acknowledgements.} This work was partly supported by the ERC Synergy Grant 610150 (imPACT). 
We thank Aniko Hannak and Abhijnan Chakraborty for inspiring discussions at the initial stage of this project.}

\bibliographystyle{ACM-Reference-Format} 
\bibliography{fair_rankings}


\begin{thebibliography}{37}


\ifx \showCODEN    \undefined \def \showCODEN     #1{\unskip}     \fi
\ifx \showDOI      \undefined \def \showDOI       #1{#1}\fi
\ifx \showISBNx    \undefined \def \showISBNx     #1{\unskip}     \fi
\ifx \showISBNxiii \undefined \def \showISBNxiii  #1{\unskip}     \fi
\ifx \showISSN     \undefined \def \showISSN      #1{\unskip}     \fi
\ifx \showLCCN     \undefined \def \showLCCN      #1{\unskip}     \fi
\ifx \shownote     \undefined \def \shownote      #1{#1}          \fi
\ifx \showarticletitle \undefined \def \showarticletitle #1{#1}   \fi
\ifx \showURL      \undefined \def \showURL       {\relax}        \fi
\providecommand\bibfield[2]{#2}
\providecommand\bibinfo[2]{#2}
\providecommand\natexlab[1]{#1}
\providecommand\showeprint[2][]{arXiv:#2}

\bibitem[\protect\citeauthoryear{Abebe, Kleinberg, and Parkes}{Abebe
  et~al\mbox{.}}{2017}]%
        {abebe2017fair}
\bibfield{author}{\bibinfo{person}{Rediet Abebe}, \bibinfo{person}{Jon
  Kleinberg}, {and} \bibinfo{person}{David~C Parkes}.}
  \bibinfo{year}{2017}\natexlab{}.
\newblock \showarticletitle{Fair division via social comparison}. In
  \bibinfo{booktitle}{\emph{AAMAS}}.
\newblock


\bibitem[\protect\citeauthoryear{Adler, Falk, Friedler, Rybeck, Scheidegger,
  Smith, and Venkatasubramanian}{Adler et~al\mbox{.}}{2016}]%
        {Adler:ICDM2016}
\bibfield{author}{\bibinfo{person}{Philip Adler}, \bibinfo{person}{Casey Falk},
  \bibinfo{person}{Sorelle Friedler}, \bibinfo{person}{Gabriel Rybeck},
  \bibinfo{person}{Carlos Scheidegger}, \bibinfo{person}{Brandon Smith}, {and}
  \bibinfo{person}{Suresh Venkatasubramanian}.}
  \bibinfo{year}{2016}\natexlab{}.
\newblock \showarticletitle{Auditing Black-Box Models for Indirect Influence}.
  In \bibinfo{booktitle}{\emph{ICDM}}.
\newblock


\bibitem[\protect\citeauthoryear{Armony and Ward}{Armony and Ward}{2010}]%
        {Armony:OR2010}
\bibfield{author}{\bibinfo{person}{Mor Armony} {and} \bibinfo{person}{Amy~R.
  Ward}.} \bibinfo{year}{2010}\natexlab{}.
\newblock \showarticletitle{Fair Dynamic Routing in Large-Scale
  Heterogeneous-Server Systems}.
\newblock \bibinfo{journal}{\emph{Operations Research}} (\bibinfo{year}{2010}).
\newblock


\bibitem[\protect\citeauthoryear{Biega, Saha~Roy, and Weikum}{Biega
  et~al\mbox{.}}{2017}]%
        {biega2017privacy}
\bibfield{author}{\bibinfo{person}{Asia~J Biega}, \bibinfo{person}{Rishiraj
  Saha~Roy}, {and} \bibinfo{person}{Gerhard Weikum}.}
  \bibinfo{year}{2017}\natexlab{}.
\newblock \showarticletitle{Privacy through Solidarity: A
  User-Utility-Preserving Framework to Counter Profiling}. In
  \bibinfo{booktitle}{\emph{SIGIR}}.
\newblock


\bibitem[\protect\citeauthoryear{Calo and Rosenblat}{Calo and
  Rosenblat}{2017}]%
        {calo2017taking}
\bibfield{author}{\bibinfo{person}{Ryan Calo} {and} \bibinfo{person}{Alex
  Rosenblat}.} \bibinfo{year}{2017}\natexlab{}.
\newblock \showarticletitle{The taking economy: Uber, information, and power}.
\newblock \bibinfo{journal}{\emph{Columbia Law Review}} (\bibinfo{year}{2017}).
\newblock


\bibitem[\protect\citeauthoryear{Celis, Straszak, and Vishnoi}{Celis
  et~al\mbox{.}}{2017}]%
        {celis2017ranking}
\bibfield{author}{\bibinfo{person}{L~Elisa Celis}, \bibinfo{person}{Damian
  Straszak}, {and} \bibinfo{person}{Nisheeth~K Vishnoi}.}
  \bibinfo{year}{2017}\natexlab{}.
\newblock \showarticletitle{Ranking with Fairness Constraints}.
\newblock \bibinfo{journal}{\emph{arXiv preprint arXiv:1704.06840}}.
\newblock


\bibitem[\protect\citeauthoryear{Chakraborty, Biega, Hannak, and
  Gummadi}{Chakraborty et~al\mbox{.}}{2017}]%
        {chakraborty2017fair}
\bibfield{author}{\bibinfo{person}{Abhijnan Chakraborty},
  \bibinfo{person}{Asia~J. Biega}, \bibinfo{person}{Aniko Hannak}, {and}
  \bibinfo{person}{Krishna~P. Gummadi}.} \bibinfo{year}{2017}\natexlab{}.
\newblock \showarticletitle{Fair Sharing for Sharing Economy Platforms}. In
  \bibinfo{booktitle}{\emph{FATREC@RecSys Workshop}}.
\newblock


\bibitem[\protect\citeauthoryear{Chuklin, Markov, and de~Rijke}{Chuklin
  et~al\mbox{.}}{2015}]%
        {chuklin2015click}
\bibfield{author}{\bibinfo{person}{Aleksandr Chuklin}, \bibinfo{person}{Ilya
  Markov}, {and} \bibinfo{person}{Maarten de Rijke}.}
  \bibinfo{year}{2015}\natexlab{}.
\newblock \showarticletitle{Click Models for Web Search}. In
  \bibinfo{booktitle}{\emph{Morgan {\&} Claypool}}.
\newblock


\bibitem[\protect\citeauthoryear{Conference}{Conference}{[n. d.]}]%
        {FATconference}
\bibfield{author}{\bibinfo{person}{FAT Conference}.} \bibinfo{year}{[n.
  d.]}\natexlab{}.
\newblock \showarticletitle{Conference on Fairness, Accountability, and
  Transparency (FAT*)}. In \bibinfo{booktitle}{\emph{\small\tt
  http://fatconference.org/resources.html}}.
\newblock


\bibitem[\protect\citeauthoryear{Craswell, Zoeter, Taylor, and Ramsey}{Craswell
  et~al\mbox{.}}{2008}]%
        {craswell2008experimental}
\bibfield{author}{\bibinfo{person}{Nick Craswell}, \bibinfo{person}{Onno
  Zoeter}, \bibinfo{person}{Michael Taylor}, {and} \bibinfo{person}{Bill
  Ramsey}.} \bibinfo{year}{2008}\natexlab{}.
\newblock \showarticletitle{An experimental comparison of click position-bias
  models}. In \bibinfo{booktitle}{\emph{WSDM}}.
\newblock


\bibitem[\protect\citeauthoryear{Dupret and Piwowarski}{Dupret and
  Piwowarski}{2008}]%
        {DBLP:conf/sigir/DupretP08}
\bibfield{author}{\bibinfo{person}{Georges Dupret} {and}
  \bibinfo{person}{Benjamin Piwowarski}.} \bibinfo{year}{2008}\natexlab{}.
\newblock \showarticletitle{A user browsing model to predict search engine
  click data from past observations}. In \bibinfo{booktitle}{\emph{SIGIR}}.
\newblock


\bibitem[\protect\citeauthoryear{Dwork, Hardt, Pitassi, Reingold, and
  Zemel}{Dwork et~al\mbox{.}}{2012}]%
        {dwork2012fairness}
\bibfield{author}{\bibinfo{person}{Cynthia Dwork}, \bibinfo{person}{Moritz
  Hardt}, \bibinfo{person}{Toniann Pitassi}, \bibinfo{person}{Omer Reingold},
  {and} \bibinfo{person}{Richard Zemel}.} \bibinfo{year}{2012}\natexlab{}.
\newblock \showarticletitle{Fairness through awareness}. In
  \bibinfo{booktitle}{\emph{ITCS}}.
\newblock


\bibitem[\protect\citeauthoryear{Edelman, Ostrovsky, and Schwarz}{Edelman
  et~al\mbox{.}}{2007}]%
        {edelman2007internet}
\bibfield{author}{\bibinfo{person}{Benjamin Edelman}, \bibinfo{person}{Michael
  Ostrovsky}, {and} \bibinfo{person}{Michael Schwarz}.}
  \bibinfo{year}{2007}\natexlab{}.
\newblock \showarticletitle{Internet advertising and the generalized
  second-price auction: Selling billions of dollars worth of keywords}.
\newblock \bibinfo{journal}{\emph{American economic review}}
  (\bibinfo{year}{2007}).
\newblock


\bibitem[\protect\citeauthoryear{Feldman, Friedler, Moeller, Scheidegger, and
  Venkatasubramanian}{Feldman et~al\mbox{.}}{2015}]%
        {Feldman:KDD2015}
\bibfield{author}{\bibinfo{person}{Michael Feldman}, \bibinfo{person}{Sorelle
  Friedler}, \bibinfo{person}{John Moeller}, \bibinfo{person}{Carlos
  Scheidegger}, {and} \bibinfo{person}{Suresh Venkatasubramanian}.}
  \bibinfo{year}{2015}\natexlab{}.
\newblock \showarticletitle{Certifying and Removing Disparate Impact}. In
  \bibinfo{booktitle}{\emph{KDD}}.
\newblock


\bibitem[\protect\citeauthoryear{Ghodsi, Sekar, Zaharia, and Stoica}{Ghodsi
  et~al\mbox{.}}{2012}]%
        {ghodsi2012multi}
\bibfield{author}{\bibinfo{person}{Ali Ghodsi}, \bibinfo{person}{Vyas Sekar},
  \bibinfo{person}{Matei Zaharia}, {and} \bibinfo{person}{Ion Stoica}.}
  \bibinfo{year}{2012}\natexlab{}.
\newblock \showarticletitle{Multi-resource fair queueing for packet
  processing}. In \bibinfo{booktitle}{\emph{SIGCOMM}}.
\newblock


\bibitem[\protect\citeauthoryear{Ghodsi, Zaharia, Hindman, Konwinski, Shenker,
  and Stoica}{Ghodsi et~al\mbox{.}}{2011}]%
        {Ghodsi:NSDI2011}
\bibfield{author}{\bibinfo{person}{Ali Ghodsi}, \bibinfo{person}{Matei
  Zaharia}, \bibinfo{person}{Benjamin Hindman}, \bibinfo{person}{Andy
  Konwinski}, \bibinfo{person}{Scott Shenker}, {and} \bibinfo{person}{Ion
  Stoica}.} \bibinfo{year}{2011}\natexlab{}.
\newblock \showarticletitle{Dominant resource fairness: Fair allocation of
  multiple resource types}. In \bibinfo{booktitle}{\emph{NSDI}}.
\newblock


\bibitem[\protect\citeauthoryear{Greenberg}{Greenberg}{1987}]%
        {greenberg1987taxonomy}
\bibfield{author}{\bibinfo{person}{Jerald Greenberg}.}
  \bibinfo{year}{1987}\natexlab{}.
\newblock \showarticletitle{A taxonomy of organizational justice theories}.
\newblock \bibinfo{journal}{\emph{Academy of Management review}}
  (\bibinfo{year}{1987}).
\newblock


\bibitem[\protect\citeauthoryear{Grgic-Hlaca, Zafar, Gummadi, and
  Weller}{Grgic-Hlaca et~al\mbox{.}}{2018}]%
        {grgic2018beyond}
\bibfield{author}{\bibinfo{person}{Nina Grgic-Hlaca},
  \bibinfo{person}{Muhammad~Bilal Zafar}, \bibinfo{person}{Krishna~P Gummadi},
  {and} \bibinfo{person}{Adrian Weller}.} \bibinfo{year}{2018}\natexlab{}.
\newblock \showarticletitle{Beyond Distributive Fairness in Algorithmic
  Decision Making: Feature Selection for Procedurally Fair Learning}. In
  \bibinfo{booktitle}{\emph{AAAI}}.
\newblock


\bibitem[\protect\citeauthoryear{Guo, Liu, and Wang}{Guo et~al\mbox{.}}{2009}]%
        {DBLP:conf/wsdm/GuoLW09}
\bibfield{author}{\bibinfo{person}{Fan Guo}, \bibinfo{person}{Chao Liu}, {and}
  \bibinfo{person}{Yi~Min Wang}.} \bibinfo{year}{2009}\natexlab{}.
\newblock \showarticletitle{Efficient multiple-click models in web search}. In
  \bibinfo{booktitle}{\emph{WSDM}}.
\newblock


\bibitem[\protect\citeauthoryear{Hardt, Price, and Srebro}{Hardt
  et~al\mbox{.}}{2016}]%
        {DBLP:conf/nips/HardtPNS16}
\bibfield{author}{\bibinfo{person}{Moritz Hardt}, \bibinfo{person}{Eric Price},
  {and} \bibinfo{person}{Nati Srebro}.} \bibinfo{year}{2016}\natexlab{}.
\newblock \showarticletitle{Equality of Opportunity in Supervised Learning}. In
  \bibinfo{booktitle}{\emph{NIPS}}.
\newblock


\bibitem[\protect\citeauthoryear{Joachims and Radlinski}{Joachims and
  Radlinski}{2007}]%
        {JoachimsRadlinski2007}
\bibfield{author}{\bibinfo{person}{Thorsten Joachims} {and}
  \bibinfo{person}{Filip Radlinski}.} \bibinfo{year}{2007}\natexlab{}.
\newblock \showarticletitle{Search Engines that Learn from Implicit Feedback}.
\newblock \bibinfo{journal}{\emph{{IEEE} Computer}} (\bibinfo{year}{2007}).
\newblock


\bibitem[\protect\citeauthoryear{Joachims, Swaminathan, and Schnabel}{Joachims
  et~al\mbox{.}}{2017}]%
        {DBLP:conf/wsdm/JoachimsSS17}
\bibfield{author}{\bibinfo{person}{Thorsten Joachims}, \bibinfo{person}{Adith
  Swaminathan}, {and} \bibinfo{person}{Tobias Schnabel}.}
  \bibinfo{year}{2017}\natexlab{}.
\newblock \showarticletitle{Unbiased Learning-to-Rank with Biased Feedback}. In
  \bibinfo{booktitle}{\emph{WSDM}}.
\newblock


\bibitem[\protect\citeauthoryear{Kamishima, Akaho, Asoh, and Sakuma}{Kamishima
  et~al\mbox{.}}{2012}]%
        {DBLP:conf/pkdd/KamishimaAAS12}
\bibfield{author}{\bibinfo{person}{Toshihiro Kamishima},
  \bibinfo{person}{Shotaro Akaho}, \bibinfo{person}{Hideki Asoh}, {and}
  \bibinfo{person}{Jun Sakuma}.} \bibinfo{year}{2012}\natexlab{}.
\newblock \showarticletitle{Fairness-Aware Classifier with Prejudice Remover
  Regularizer}. In \bibinfo{booktitle}{\emph{ECML/PKDD}}.
\newblock


\bibitem[\protect\citeauthoryear{Kearns, Roth, and Wu}{Kearns
  et~al\mbox{.}}{2017}]%
        {Kearns:ICML2017}
\bibfield{author}{\bibinfo{person}{Michael Kearns}, \bibinfo{person}{Aaron
  Roth}, {and} \bibinfo{person}{Zhiwei~Steven Wu}.}
  \bibinfo{year}{2017}\natexlab{}.
\newblock \showarticletitle{Meritocratic Fairness for Cross-Population
  Selection}. In \bibinfo{booktitle}{\emph{ICML}}.
\newblock


\bibitem[\protect\citeauthoryear{Kleinberg, Lakkaraju, Leskovec, Ludwig, and
  Mullainathan}{Kleinberg et~al\mbox{.}}{2017}]%
        {kleinberg2017human}
\bibfield{author}{\bibinfo{person}{Jon Kleinberg}, \bibinfo{person}{Himabindu
  Lakkaraju}, \bibinfo{person}{Jure Leskovec}, \bibinfo{person}{Jens Ludwig},
  {and} \bibinfo{person}{Sendhil Mullainathan}.}
  \bibinfo{year}{2017}\natexlab{}.
\newblock \showarticletitle{Human decisions and machine predictions}.
\newblock \bibinfo{journal}{\emph{The Quarterly Journal of Economics}}
  (\bibinfo{year}{2017}).
\newblock


\bibitem[\protect\citeauthoryear{Levy and Barocas}{Levy and Barocas}{2018}]%
        {levy2017designing}
\bibfield{author}{\bibinfo{person}{Karen Levy} {and} \bibinfo{person}{Solon
  Barocas}.} \bibinfo{year}{2018}\natexlab{}.
\newblock \showarticletitle{Designing Against Discrimination in Online
  Markets}.
\newblock \bibinfo{journal}{\emph{Berkeley Technology Law Journal}}
  (\bibinfo{year}{2018}).
\newblock


\bibitem[\protect\citeauthoryear{Mehrotra, Anderson, Diaz, Sharma, Wallach, and
  Yilmaz}{Mehrotra et~al\mbox{.}}{2017}]%
        {Mehrotra:WWW2017}
\bibfield{author}{\bibinfo{person}{Rishabh Mehrotra}, \bibinfo{person}{Ashton
  Anderson}, \bibinfo{person}{Fernando Diaz}, \bibinfo{person}{Amit Sharma},
  \bibinfo{person}{Hanna Wallach}, {and} \bibinfo{person}{Emine Yilmaz}.}
  \bibinfo{year}{2017}\natexlab{}.
\newblock \showarticletitle{Auditing Search Engines for Differential
  Satisfaction Across Demographics}. In \bibinfo{booktitle}{\emph{WWW}}.
\newblock


\bibitem[\protect\citeauthoryear{Pedreschi, Ruggieri, and Turini}{Pedreschi
  et~al\mbox{.}}{2008}]%
        {DBLP:conf/kdd/PedreschiRT08}
\bibfield{author}{\bibinfo{person}{Dino Pedreschi}, \bibinfo{person}{Salvatore
  Ruggieri}, {and} \bibinfo{person}{Franco Turini}.}
  \bibinfo{year}{2008}\natexlab{}.
\newblock \showarticletitle{Discrimination-aware data mining}. In
  \bibinfo{booktitle}{\emph{KDD}}.
\newblock


\bibitem[\protect\citeauthoryear{Romei and Ruggieri}{Romei and
  Ruggieri}{2014}]%
        {DBLP:journals/ker/RomeiR14}
\bibfield{author}{\bibinfo{person}{Andrea Romei} {and}
  \bibinfo{person}{Salvatore Ruggieri}.} \bibinfo{year}{2014}\natexlab{}.
\newblock \showarticletitle{A multidisciplinary survey on discrimination
  analysis}.
\newblock \bibinfo{journal}{\emph{Knowledge Eng. Review}}
  (\bibinfo{year}{2014}).
\newblock


\bibitem[\protect\citeauthoryear{Singh and Joachims}{Singh and
  Joachims}{2018}]%
        {SinghJoachims2017}
\bibfield{author}{\bibinfo{person}{Ashudeep Singh} {and}
  \bibinfo{person}{Thorsten Joachims}.} \bibinfo{year}{2018}\natexlab{}.
\newblock \showarticletitle{Fairness of Exposure in Rankings}.
\newblock \bibinfo{journal}{\emph{arXiv preprint arXiv:1802.07281}}.
\newblock


\bibitem[\protect\citeauthoryear{Walster, Berscheid, and Walster}{Walster
  et~al\mbox{.}}{1973}]%
        {walster1973new}
\bibfield{author}{\bibinfo{person}{Elaine Walster}, \bibinfo{person}{Ellen
  Berscheid}, {and} \bibinfo{person}{G~William Walster}.}
  \bibinfo{year}{1973}\natexlab{}.
\newblock \showarticletitle{New directions in equity research.}
\newblock \bibinfo{journal}{\emph{Journal of personality and social
  psychology}} (\bibinfo{year}{1973}).
\newblock


\bibitem[\protect\citeauthoryear{Wang, Bendersky, Metzler, and Najork}{Wang
  et~al\mbox{.}}{2016}]%
        {DBLP:conf/sigir/WangBMN16}
\bibfield{author}{\bibinfo{person}{Xuanhui Wang}, \bibinfo{person}{Michael
  Bendersky}, \bibinfo{person}{Donald Metzler}, {and} \bibinfo{person}{Marc
  Najork}.} \bibinfo{year}{2016}\natexlab{}.
\newblock \showarticletitle{Learning to Rank with Selection Bias in Personal
  Search}. In \bibinfo{booktitle}{\emph{SIGIR}}.
\newblock


\bibitem[\protect\citeauthoryear{Yaari and Bar-Hillel}{Yaari and
  Bar-Hillel}{1984}]%
        {yaari1984dividing}
\bibfield{author}{\bibinfo{person}{Menahem~E Yaari} {and} \bibinfo{person}{Maya
  Bar-Hillel}.} \bibinfo{year}{1984}\natexlab{}.
\newblock \showarticletitle{On dividing justly}.
\newblock \bibinfo{journal}{\emph{Social choice and welfare}}
  (\bibinfo{year}{1984}).
\newblock


\bibitem[\protect\citeauthoryear{Yang and Stoyanovich}{Yang and
  Stoyanovich}{2007}]%
        {yang2016measuring}
\bibfield{author}{\bibinfo{person}{Ke Yang} {and} \bibinfo{person}{Julia
  Stoyanovich}.} \bibinfo{year}{2007}\natexlab{}.
\newblock \showarticletitle{Measuring fairness in ranked outputs}. In
  \bibinfo{booktitle}{\emph{SSDBM}}.
\newblock


\bibitem[\protect\citeauthoryear{Zafar, Valera, Gomez{-}Rodriguez, and
  Gummadi}{Zafar et~al\mbox{.}}{2017}]%
        {DBLP:conf/www/ZafarVGG17}
\bibfield{author}{\bibinfo{person}{Muhammad~Bilal Zafar},
  \bibinfo{person}{Isabel Valera}, \bibinfo{person}{Manuel Gomez{-}Rodriguez},
  {and} \bibinfo{person}{Krishna~P. Gummadi}.} \bibinfo{year}{2017}\natexlab{}.
\newblock \showarticletitle{Fairness Beyond Disparate Treatment {\&} Disparate
  Impact: Learning Classification without Disparate Mistreatment}. In
  \bibinfo{booktitle}{\emph{WWW}}.
\newblock


\bibitem[\protect\citeauthoryear{Zehlike, Bonchi, Castillo, Hajian, Megahed,
  and Baeza-Yates}{Zehlike et~al\mbox{.}}{2017}]%
        {zehlike2017fair}
\bibfield{author}{\bibinfo{person}{Meike Zehlike}, \bibinfo{person}{Francesco
  Bonchi}, \bibinfo{person}{Carlos Castillo}, \bibinfo{person}{Sara Hajian},
  \bibinfo{person}{Mohamed Megahed}, {and} \bibinfo{person}{Ricardo
  Baeza-Yates}.} \bibinfo{year}{2017}\natexlab{}.
\newblock \showarticletitle{{FA*IR}: A fair top-k ranking algorithm}. In
  \bibinfo{booktitle}{\emph{CIKM}}.
\newblock


\bibitem[\protect\citeauthoryear{Zemel, Wu, Swersky, Pitassi, and Dwork}{Zemel
  et~al\mbox{.}}{2013}]%
        {DBLP:conf/icml/ZemelWSPD13}
\bibfield{author}{\bibinfo{person}{Richard~S. Zemel}, \bibinfo{person}{Yu Wu},
  \bibinfo{person}{Kevin Swersky}, \bibinfo{person}{Toniann Pitassi}, {and}
  \bibinfo{person}{Cynthia Dwork}.} \bibinfo{year}{2013}\natexlab{}.
\newblock \showarticletitle{Learning Fair Representations}. In
  \bibinfo{booktitle}{\emph{ICML}}.
\newblock


\end{thebibliography}

\end{document}